\documentclass{article}

\usepackage{amsfonts}
\usepackage{amssymb}
\usepackage{amstext}

\newcommand{\slim}{\mathop{\mbox{\rm s-lim}}}

\newcommand\R{{\mathbb R}}
\newcommand\C{{\mathbb C}}
\newcommand\Z{{\mathbb Z}}

\renewcommand\H{{\mathcal H}}
\newcommand\M{{\mathcal M}}
\newcommand\NN{{\mathcal N}}
\newcommand\RR{{\mathcal R}}

\renewcommand\l{\lambda}
\newcommand\om{\omega}

\newcommand\ds{\displaystyle}

\begin{document}

\title{Selfadjoint time operators and \\ invariant subspaces}

\author{Fernando G\'omez$^1$}

\maketitle

\begin{center}
\small
$^1$Dpto. de An\'alisis Matem\'atico, Universidad de Valladolid,\\
Facultad de Ciencias, Prado de la Magdalena, s.n.,\\
47005 Valladolid, Spain.\\
email: {\tt fgcubill@am.uva.es}
\end{center}

\begin{abstract}
For classical dynamical systems time operators are introduced as selfadjoint operators satisfying the so called weak Weyl relation (WWR) with the unitary groups of time evolution.
Dynamical systems with time operators are intrinsically irreversible because they admit Lyapounov operators as functions of the time operator.
For quantum systems  selfadjoint time operators are defined in the same way or as dilations of symmetric ones dealing with the time-energy uncertainty relation, times of occurrence or survival probabilities.  
This work tackles the question of the existence of selfadjoint time operators on the basis of the Halmos-Helson theory of invariant subspaces, the Sz.-Nagy-Foia{\c s} dilation theory and the Misra-Prigogine-Courbage theory of irreversibility.
It is shown that the existence of a selfadjoint time operator for a unitary evolution is equivalent to the intrinsic irreversibility of the evolution plus the existence of a simply invariant subspace or a rigid operator-valued function for its Sz.-Nagy-Foia{\c s} functional model. 
An extensive set of equivalent conditions to the existence of selfadjoint time operators can be obtained from these results. Such conditions are written in terms of Schr\"odinger couples, the Weyl commutation relation, incoming and outgoing subspaces, innovation processes, Lax-Phillips scattering processes, and translation and spectral representations. 
As an example, the selfadjoint extension of the quantum Aharonov-Bohm time-of-arrival operator is studied. 
\end{abstract}

{\small


}

\section{Introduction}

The conventional topological approach to the study of classical dynamical systems is based on trajectories in the phase space $\Omega$ describing the point dynamics by a family $S_t$ of endomorphisms or automorphisms of $\Omega$, namely, the time evolution
$$
\om_0\mapsto \om_t:=S_t\om_0
$$
of single points $\om_0\in\Omega$, where $t\in\R$ or $t\in\R^+:=[0,\infty)$ for flows and $t\in\Z$ or $t\in\Z^+:=\{n\in\Z:n\geq0\}$ for cascades. 
In the probabilistic approach, extensively used in statistical mechanics and ergodic theory, trajectories are replaced by the study of the corresponding Koopman and/or Frobenius--Perron operators \cite{LM94}, which describe, respectively, the evolution of the observables and the probability densities of the system.
In the Hilbert space $L^2=L^2(\Omega,{\mathcal A},\mu)$ of square integrable functions on $\Omega$ --with respect to the reference $\sigma$-algebra ${\mathcal A}$ and measure $\mu$-- the Koopman operator $V_t$ and its $L^2$-adjoint, the Frobenius--Perron operator $U_t$, are defined as
$$
V_tf(\om):= f(S_t\om)\,,\qquad  (U_t\rho,f)=(\rho,V_tf)\,,
$$
where $(\rho,f)=\int_\Omega \rho(\om)\,f(\om)\,d\mu(\om)$ is the expectation value of the observable $f$ in the density $\rho$.

In the context of statistical mechanics, Misra, Prigogine and Courbage (MPC) \cite{M78}-\cite{MPC79c} \cite{CM80} \cite{GMC81} introduced an exact theory of irreversibility showing that for some reversible dynamical systems, described by unitary groups $\{U_t\}$, it is possible to construct  non-unitary similarity transformations $\Lambda$ (see Section \ref{sschs}) which convert the groups $\{U_t\}$ into contraction semigroups $\{W_t\}$ corresponding to irreversible evolutions.
It is usual to call such unitary evolutions ``intrinsically irreversible" because they admit Lyapounov operators.
The construction of $\Lambda$ was first done for a special dynamical system, the baker transformation \cite{MPC79c},  and soon generalized to Bernoulli systems \cite{CM80} and  Kolmogorov systems \cite{GMC81}.

Following a suggestion by Misra \cite{M78}, the unitary evolutions $\{U_t\}$ for which $\Lambda$ transformations have been constructed have been qualified by the existence of an {\it (internal) time operator}, a self-adjoint operator $T$ that satisfies the {\it weak Weyl relation} (WWR): for every $\rho\in\text{Dom}(T)$ and $t\in\R$ (or $t\in\Z$), one has $U_t\rho\in\text{Dom}(T)$ and 
$$
U_{-t} T U_t\,\rho=(T+tI)\,\rho
$$
($\text{Dom}(T)$ denotes the domain of $T$). This relation has perhaps the most direct physical interpretation: the  time operator $T$ allows the attribution of an {\it average age} $(\rho,T\rho)$ to the states $\rho$ which keeps step with the external clock time $t$ for the evolved state $U_t\rho$:
$$
(U_t\rho,TU_t\rho)=(\rho,T\rho)+t\,.
$$
The transformation $\Lambda$ and the Lyapounov operator are then operator functions of the time operator $T$.

Paying attention to flows only, it has been shown \cite{Mi01,Ga02} that not only the expectation value but the entire probability distribution of ages is shifted by the external time, i.e. the  time operator $T$ satisfy the {\it generalized weak Weyl relation} (GWWR): for any $t\in\R$, 
$$
U_{-t} E_T(\cdot)U_{t}=E_T(\cdot-t)\,,
$$
where $E_T(\cdot)$ denotes the spectral measure of $T$. In other words, the time operator and the selfadjoint generator $A$ of the time evolution $\{U_t=e^{itA}\}_{t\in\R}$ form a system of imprimitivities based on the real line $\R$ (see \cite{Mack78}).
The WWR is equivalent to the GWWR as well as to the usual {\it Weyl relation} (WR): for all $s,t\in\R$,
$$
e^{itA}e^{isT}=e^{its}e^{isT}e^{itA}\,.
$$
Moreover, in this case, the WWR --GWWR or WR-- implies the {\it canonical commutation relation} (CCR): for every $\rho\in \text{Dom}(TA)\cap\text{Dom}(AT)$,
$$
[T,A]\rho:=TA\rho-AT\rho=i\rho\,,
$$
but the converse is not true in general (see also \cite[Sect.VIII.5]{RS72}).

That all solutions $(T,A)$ of the WR are unitarily equivalent to the momentum-position Schr\"odinger couple $(P,Q)$ or a direct sum of such couples was proved by von Neumann \cite{VN31}. This implies in particular that $T$ and $A$ have absolutely continuous and uniform spectra spanning the entire real line.

Further work has been done extending the concept of  time operator in different contexts. Time operators have been introduced for Liouville quantum dynamics \cite{MPC79b,Courb80,Cour01}, relativistic fields \cite{AnMis92}, unilateral shifts \cite{ASS99}, diffusions \cite{APSS00}, semigroups with normal generators \cite{SS03} or dilations of Markov processes \cite{AS03}. 
Also the connections of the  time operators with innovation processes \cite{WOLD54} and Lax-Phillips scattering processes \cite{LP67} have been repeatedly emphasized  as in \cite{GM76,AG00,A01,SA03} and \cite{M87,AnMis92,ASS99,A01}, respectively.

In standard quantum mechanics {\it time} is a external variable $t$ independent from the dynamics of any given system, but it acquires dynamical significance in questions involving the occurrence of an event or the time-energy uncertainty relation (see \cite{Ga02} and references therein). 
When this uncertainty relation is considered the time operator $T$ is usually defined to satisfy the CCR with the Hamiltonian $H$ (the generator of the time evolution).
In the attempt to define quantum mechanical observables describing the time of occurrence of an event it is natural to start from the WWR.
Since the Hamiltonian $H$ generally possesses a semibounded spectrum and  a selfadjoint time operator cannot exist, symmetric (but not selfadjoint) time operators have been extensively studied \cite{MSE02}. A particular example is (an extension $\tilde T_0$ of) the so called {\it Aharonov-Bohm time operator} \cite{AB61}
$$
T_0=\frac12\left(QP^{-1}+P^{-1}Q\right)\,,
$$ 
which obeys the WWR with respect to the one-dimensional free Hamiltonian $H_0=P^2/2$. 
Miyamoto \cite{Mi01} has shown the deep connection between such symmetric time operators and the survival probability.
Dorfmeister and Dorfmeister \cite{Dorf84} and Schm\"udgen \cite{Sch83} gave structure theorems for the pairs $(T,H)$ consisting of a symmetric operator $T$ and an essentially  selfadjoint operator $H$ defined on a dense invariant domain $\mathcal D$ and satisfying the WWR on $\mathcal D$.
These theorems show that these generalized pairs are not very different from direct sums of Schr\"odinger couples $(P,Q)$.  
In particular, the spectrum of $H$ is absolutely continuous, i.e., the system consists of only scattering states, and $T$ has no point spectrum.

It is also possible to construct positive operator valued (but not projection valued) measures satisfying the GWWR . Such measures have the same statistical interpretation as the projection valued ones and will hence be considered observables \cite{Dav76}.
It is clear that the GWWR implies the WWR for the operator $T=\int x\,E(dx)$ defined in a suitable domain. That every symmetric operator $T$ satisfying the WWR arises in this way was proved by Werner \cite{Wer90}. Therefore, in this case, one has also the equivalence of the WWR and the GWWR.
This entails the existence of a selfadjoint extension of $T$ and corresponding projection values measure to a larger Hilbert space satisfying the analogous commutation relation with a suitable extended unitary group \cite{Wer90}.

This work tackles the question of the existence of selfadjoint  time operators on the basis of the Halmos-Helson theory of invariant subspaces \cite{HALMOS61,HELSON64}, the Sz.-Nagy-Foia{\c s} dilation theory \cite{NAGY-FOIAS} and the MPC theory of irreversibility \cite{M78}-\cite{MPC79c}.
It is shown that the existence of a selfadjoint  time operator for a unitary flow or cascade $\{U_t\}$ is equivalent to the intrinsic irreversibility of $\{U_t\}$ plus the existence of a simply invariant subspace or a rigid operator-valued function for the Sz.-Nagy-Foia{\c s} functional model of $\{U_t\}$. 
From these results an extensive set of equivalent conditions to the existence of selfadjoint  time operators can be deduced as direct corollaries. Such conditions are written in terms of Schr\"odinger couples, the Weyl commutation relation, incoming and outgoing subspaces, innovation processes, Lax-Phillips scattering processes, and translation and spectral representations. 
Moreover, this approach permits to handle flows and cascades on the same foot. 
As an example, it is also shown how  the Aharonov-Bohm pair $(\tilde T_0,H_0)$ may be obtained by projection from a twofold Schr\"odinger couple $(P,Q)$. 

The paper is organized as follows.
Section \ref{sis} includes a review of the theory of invariant subspaces developed by Beurling \cite{BEUR49}, Halmos \cite{HALMOS61}, Helson \cite{HELSON64,HL61}, Lax \cite{LAX59,LAX61} and Srinivasan \cite{SRI63, SRI64}, among others --see \cite{RadR03} and references therein--. This theory describes the doubly and simply invariant subspaces of certain spaces of vector-valued functions over the complex unit circle $C$ in terms of certain operator-valued functions, the so called range and rigid functions. The picture is completed in Section \ref{sfm}.
Some connections between the MPC theory of irreversibility and the Sz.-Nagy-Foia{\c s} dilation theory obtained in a previous work by the author \cite{G} are presented in Section \ref{sschs}.
In section \ref{sfm} two functional models for intrinsically irreversible unitary evolutions are introduced. These functional models are based on the Sz.-Nagy-Foia{\c s} functional calculus and  given in terms of vector-valued functions on the unit circle $C$ and the real line $\R$. The first one over $C$ is well-adapted for cascades and the second one over $\R$ for flows. The main concepts and results about invariant subspaces introduced in Section \ref{sis} over $C$ are transfered into the functional calculus over $\R$ and a key result in our development is proved: the spaces where the functional models live are doubly invariant subspaces and, moreover, the corresponding range functions are those given by the Sz.-Nagy-Foia{\c s} calculus.
Section \ref{sito} tackle the question of the existence of selfadjoint  time operators.
The extensive set of equivalent conditions for both flows and cascades are presented and proved there.
Finally, the study of the Aharonov--Bohm time operator is presented in Section \ref{sabto}.

\section{Invariant subspaces}\label{sis}

Let $C$ be the unit circle and let $D$ be the open unit disc of the complex plane $\mathbb C$,
$$
C:=\{\omega\in\C:|\omega|=1\},\qquad D:=\{\lambda\in\C:|\lambda|<1\}.
$$
In $C$ interpret measurability in the sense of Borel and consider the normalized Lebesgue measure $d\omega/(2\pi)$.
Given a separable Hilbert space $\H$, let $L^2(\H)$ denote the set of all measurable functions $v:C\to \H$ such that ${1\over 2\pi}\int_C ||v(\omega)||^2_\H\,d\omega<\infty$ (modulo sets of measure zero); measurability here can be interpreted either strongly or weakly, which amounts to the same due to the separability of $\H$. The functions in $L^2(\H)$ constitute a Hilbert space with pointwise definition of linear operations and inner product given by 
$$
(u,v):={1\over 2\pi}\int_C \big(u(\omega),v(\omega)\big)_\H\,d\omega\,,\qquad u,v\in L^2(\H)\,.
$$
Let us denote by $H^2(\H)$ the Hardy class of functions
$$
\tilde u(\lambda)=\sum_{k=0}^\infty \lambda^k a_k,\qquad (\lambda\in D,\,a_k\in\H),
$$
with values in $\H$, holomorphic on $D$, and such that
${1\over 2\pi}\int_C ||\tilde u(r\omega)||^2_\H\,d\omega$, ($0\leq r<1$),
has a bound independent of $r$ or, equivalently, such that $\sum ||a_k||^2_\H<\infty$.
For each function $\tilde u\in H^2(\H)$ the non-tangential limit in strong sense 
$$
\slim_{\lambda\to\omega} \tilde u(\lambda)=\sum_{k=0}^\infty \omega^k a_k=: u(\omega)
$$
exist for almost all $\omega\in C$. The functions $\tilde u(\lambda)$ and $ u(\omega)$ determine each other (they are connected by Poisson formula), so that we can identify $H^2(\H)$ with a subspace of $L^2(\H)$, say $L^2_+(\H)$, thus providing $H^2(\H)$ with the Hilbert space structure of $L^2_+(\H)$ and embedding it in $L^2(\H)$ as a subspace.  
Fixing some orthonormal basis $\{u_1,u_2,\cdots\}$ for $\H$, each element $v$ of $L^2(\H)$ may be expressed in terms of its relative coordinate functions $f_j$,
$$
v(\omega)=\sum_{j}f_j(\omega)\cdot u_j\,,
$$
where the functions $f_j$ are in $L^2:=L^2(\C)$ and $||v||^2=\sum_{j} ||f_j||^2$. 
A function $v$ in $L^2(\H)$ belongs to $H^2(\H)$ iff its coordinate functions $f_j$ are all in the Hardy space $H^2:=H^2(\C)$.
Elementary properties of vector and operator valued functions are given in Hille and Phillips \cite[Chapter III]{HP74}; see also \cite{HALMOS61}, \cite[Lecture VI]{HELSON64} and \cite[Sect.V.1]{NAGY-FOIAS}.

A {\it range function} $J=J(\omega)$ is a function on the unit circle $C$ taking values in the family of closed subspaces of $\H$. $J$ is measurable if the orthogonal projection $P(\omega)$ on $J(\omega)$ is weakly measurable in the operator sense, i.e. $(P(\omega)h,g)$ is a measurable scalar function for each $h,g\in\H$. Range functions which are equal almost everywhere (a.e.) on $C$ are identified.
For each measurable range function $J$, $\M_J$ denotes the set of all functions $v$ in $L^2(\H)$ such that $v(\omega)$ lies in $J(\omega)$ a.e..

A closed subspace $\M$ of $L^2(\H)$ is called {\it invariant} if $\chi\cdot v\in\M$ for every $v\in\M$, where $\chi(\omega)=\omega$.
$\M$ is called {\it doubly invariant} if $\chi\cdot v$ and $\chi^{-1}\cdot v$ belong to $\M$ for each $v\in\M$.
$\M$ is called {\it simply invariant} if it is invariant but not doubly invariant.

In the scalar case the doubly invariant subspaces were described by Wiener \cite[Th.2]{HELSON64}: 
For each measurable subset $E$ of the circle $C$, let $\M_E$ denote the set of functions in $L^2$ which vanish a.e. on the complement of $E$. {\it $\M_E$ is doubly invariant and every doubly invariant subspace is of this form.} The generalization of Wiener theorem is implicitly contained in the work of Lax \cite{LAX61} and made clear by T.P. Srinivasan \cite{SRI63, SRI64}:
{\it The doubly invariant subspaces of $L^2(\H)$ are the subspaces $\M_J$, where $J$ is a measurable range function.} The correspondence between $J$ and $\M_J$ is one-to-one, under the convention that range functions are identified if they are equal a.e..
When $\H$ is one-dimensional its only subspaces are the trivial ones $\{0\}$ and $\H$ itself; in this case a range function is known by the set of points where it is equal to $\H$, so that we have just the Wiener theorem.

In what follows by the {\it range} of a doubly invariant subspace $\M_J$ we mean that range function $J$. This definition can be extended to an arbitrary set of functions: the {\it range} of a set of vector functions is the range of the smallest doubly invariant subspace containing all the functions.

The problem of describing the simply invariant subspaces in the scalar case was solved by Beurling \cite{BEUR49}, anyway for subspaces which are contained in $H^2$. A direct proof which establishes the result for subspaces of $L^2$ was given by Helson and Lowdenslager \cite{HL61}: 
{\it The simply invariant subspaces $\M$ of $L^2$ are of the form $q\cdot H^2$, where $q$ is an arbitrary measurable function such that $|q(\omega)|=1$ a.e.. The function $q$ is determined by $\M$ up to a constant factor and when $\M$ is contained in $H^2$ then $q$ belongs to $H^2$} \cite[Th.3]{HELSON64}. The analytic functions of modulus one a.e. on $C$ were called {\it inner functions} by Beurling. Each inner function $q$ is a product $c\cdot B\cdot S$, where $c$ is a complex constant of modulus one, $B$ a Blaschke function and $S$ a singular function \cite{HOFF62}. 

Let ${\mathcal L}(\H)$ be the set of bounded linear operators on $\H$. A holomorphic ${\mathcal L}(\H)$-valued function $U$ on $D$ is called {\it inner} if the operator $[Uv](\lambda):=U(\lambda)\,v(\lambda)$, ($v\in H^2(\H)$), is a partial isometry on $H^2(\H)$. In such case there exists a subspace $\M$ of $\H$ such that
$$
\{v\in H^2(\H): ||Uv||=||v||\}= H^2(\M)
$$
and $Uv=0$ for all $v\in H^2(\M^\perp)$.
Moreover, the values of the non-tangential boundary function of $U$ are partial isometries on $\H$ with initial space $\M$ a.e. on $C$. See \cite[Sect.5.3]{RR85}. 
Halmos \cite{HALMOS61} called {\it rigid operator function} to every weakly measurable ${\mathcal L}(\H)$-valued function $U$ on $C$ such that $U(\omega)$ is for almost all $\omega\in C$ a partial isometry on ${\mathcal H}$ with the same initial space.
Lax generalization \cite{LAX59} of Beurling theorem states that {\it every invariant subspace contained in $H^2(\H)$ is of the form $U\,H^2(\H)$ for some ${\mathcal L}(\H)$-valued inner function $U$ on $D$.}
The general form of the simply invariant subspaces of $L^2(\H)$ given by Helson \cite[Th.9]{HELSON64} on the basis of the work of Halmos \cite{HALMOS61} is as follows: 
{\it Each simply invariant subspace $\M$ of $L^2(\H)$ has the form
\begin{equation}\label{H64.6.31}
U\, H^2({\mathcal H})\oplus \M_K\,,
\end{equation}
where $K$ is a measurable range function and $U$ is a rigid ${\mathcal L}(\H)$-valued function with range $J$ orthogonal to $K$ almost everywhere.}
 
Let $\M$ be a simply invariant subspace of $L^2(\H)$, let $\M_n$ be $\chi^n\cdot\M$ for each integer $n$, and define (the overbar denotes adherence)
$$
\M_\infty:=\bigcap_{n<\infty} \M_n\,,\qquad
\M_{-\infty}:=\overline{\bigcup_{n>-\infty}\M_n}\,.
$$
The subspaces $\M_\infty$ and $\M_{-\infty}$ are doubly invariant, so that they are of the form $\M_K$ and $\M_{J_0}$ for some measurable range functions $K$ and $J_0$, respectively.
The orthogonal complement ${\mathcal N}$ of $\M_\infty$ in $\M$ is a simply invariant subspace such that ${\mathcal N}_\infty=\{0\}$. The decomposition $\M={\mathcal N}\oplus\M_K$ is orthogonal in the sense of $L^2(\H)$, but further in the pointwise sense a.e.. It follows that the range $J$ of ${\mathcal N}$ is exactly the orthogonal complement of $K$ in $J_0$.

The idea of constructing {\it innovation subspaces} is due to Wold \cite{WOLD54} in connection with the prediction theory of stationary time series.
Here the innovation subspaces are defined to be ${\mathcal I}_n=\M_n\ominus\M_{n+1}$ for each integer $n$.
Evidently ${\mathcal I}_n=\chi^n\cdot{\mathcal I}_0$ and, because ${\mathcal N}_\infty=\{0\}$, we have
\begin{equation}\label{h6.34}
{\mathcal N}={\mathcal I}_0\oplus{\mathcal I}_1\oplus{\mathcal I}_2\oplus\cdots\,.
\end{equation}
Select an orthonormal basis $\{u_1,u_2,\ldots\}$ for ${\mathcal I}_0$.
For each $n$, $\{\chi^n\cdot u_j\}_j$ is an orthonormal basis for ${\mathcal I}_n$, so by (\ref{h6.34}) the set
$\{\chi^n\cdot u_j\}$, ($n\geq 0$, all $j$), is an orthonormal basis for ${\mathcal N}$.
If we allow $n$ to run over negative values as well, this set is a basis for ${\mathcal N}_{-\infty}=\M_J$.
Therefore the expansions
$$
v=\sum_{n,j} a_{nj}\chi^n\cdot u_j\,,\qquad \sum_{n,j} |a_{nj}|^2<\infty\,,
$$
cover precisely the functions $v$ in ${\mathcal N}$ when $n$ is restricted to be non-negative, and give all the functions of $\M_J$ when $n$ runs over all integers. The sum converges in the metric of $L^2(\H)$.
In terms of relative coordinate functions,
\begin{equation}\label{h6.41}
v=\sum_{j}f_j\cdot u_j\,,\qquad f_j:=\sum_{n} a_{nj}\chi^n\,,
\end{equation}
where the functions $f_j$ are in $L^2$ always and in $H^2$ if and only if $v$ belongs to ${\mathcal N}$, and they are perfectly arbitrary under the restriction $\sum_{j} ||f_j||^2=\sum_{n,j} |a_{nj}|^2<\infty$.

Now let $\H_1$ be a Hilbert space having the same dimension as ${\mathcal I}_0$, let $\{e_1,e_2,\ldots\}$ be an orthonormal basis for $\H_1$ and $U$ the operator which maps $\H_1$ isometrically on ${\mathcal I}_0$ by carrying each $e_j$ to $u_j$. This operator induces a decomposable isometry also denoted by $U$ from $L^2(\H_1)$ into $L^2(\H)$ by setting
\begin{equation}\label{h6.43}
U\sum_j f_je_j:=\sum_j f_j\cdot u_j\,.
\end{equation}
The range of $U$ is $\M_J={\mathcal N}_{-\infty}$ and when $U$ is restricted to $H^2(\H_1)$ we obtain all expansion (\ref{h6.41}) in which each $f_j$ is analytic, so that ${\mathcal N}=UH^2(\H_1)$.
The subspace ${\mathcal N}$ uniquely determines the rigid operator function $U$ to within a constant partially isometric factor on the right \cite[Th.4]{HALMOS61}.


\section{Intrinsic irreversibility of unitary evolutions}\label{sschs}

By a {\it (one-continuous-parameter) semigroup} on a separable Hilbert space $\H$ we mean a family $\{W_t\}_{t\in\R^+}\subset {\mathcal L}(\H)$ with the following properties:
\begin{itemize}
\item[(1)] 
$W_tW_s=W_{t+s}$, for $t,s\in\R^+$,
\item[(2)]
$W_0=I$,
\item[(3)]
$\ds\slim_{t\to s} W_th=W_{s}h$, for each $s\geq 0$ and $h\in \H$, i.e. 
$\ds\slim_{t\to s} W_t=W_{s}$, where $\slim$ denotes limit in strong sense in both $\H$ and ${\mathcal L}(\H)$.
\end{itemize}
A family $\{W_t\}_{t\in\R}$ is called a {\it (one-parameter) group} if it satisfies (2) as well (1) and (3) for $t,s\in\R$. Thus, from (1) and (2), $W_{-t}=W_t^{-1}$. In particular, we will be interested in unitary groups $\{U_t\}_{t\in\R}$ for which $U_{-t}=U_t^{-1}=U_t^{*}$ (the $^*$ denotes adjoint).

A {\it (one-discrete-parameter) semigroup} on $\H$ will be a family $\{W_t\}_{t\in\Z^+}\subset {\mathcal L}(\H)$ with the following properties:
\begin{itemize}
\item[(1')] 
$W_nW_m=W_{n+m}$, for $n,m\in\Z^+$,
\item[(2')]
$W_0=I$.
\end{itemize}
A family $\{W_n\}_{n\in\Z}$ is called a {\it (one-discrete-parameter) group} if it satisfies (2') as well (1') $n,m\in\Z$. From (1) and (2), $W_{-n}=W_n^{-1}$. In particular, a unitary group $\{W_n\}_{n\in\Z}$ will be of the form $\{U^n\}_{n\in\Z}$ for some unitary operator $U$.

Reversible evolutions are expressed in terms of groups of unitary operators $\{U'_t\}_{t\in\R}$ [$\{U'^n\}_{n\in\Z}$] (acting on a Hilbert space $\H'$), whereas 
irreversible dynamics are described by contraction semigroups $\{W_t\}_{t\in\R^+}$ [$\{W_n\}_{n\in\Z}$] (on a Hilbert space $\H$).
In the non-unitary transformation theory of dynamical systems, introduced as an exact theory of irreversibility by Misra, Prigogine and Courbage (MPC), an ``intrinsically irreversible" unitary group $\{U'_t\}_{t\in\R}$ [$\{U'^n\}_{n\in\Z}$] is related to a contraction semigroup $\{W_t\}_{t\in\R^+}$ [$\{W_n\}_{n\in\Z}$] through a intertwining transformation $\Lambda$:
\begin{equation}\label{Int.Relx}
W_t\Lambda=\Lambda U'_t\,,\, \text{ for } t\in\R^+,\qquad
[W_n\Lambda=\Lambda U'^n\,,\, \text{ for } n\in\Z^+,]\,,
\end{equation}
being $\Lambda$ a {\it quasi-affinity} from $\H'$ into $\H$, i.e. a linear, one-to-one and continuous transformation from $\H'$ onto a dense subspace of $\H$, so that $\Lambda^{-1}$ exists on this dense domain, but is not necessarily continuous.

A characterization of the contraction semigroups induced by the MPC theory, the structure  of the intrinsically irreversible unitary groups and a prototype for the transformations $\Lambda$ have been given in \cite{G} on the basis of the Sz.-Nagy-Foia{\c s} dilation theory and the derived functional calculus for contractions on Hilbert spaces \cite{NAGY-FOIAS}. 
The main results of \cite{G} will be presented in the sequel.

First, let us recall that, for a semigroup $\{W_t\}_{t\in\R^+}$ of contractions with infinitesimal generator $A$, the {\it cogenerator} $W$ of $\{W_t\}_{t\in\R^+}$ is given by the Cayley transform of $A$, i.e.
$$
W=(A+I)(A-I)^{-1},\qquad A=(W+I)(W-I)^{-1}.
$$
The cogenerator $W$ is then a contraction which does not have $1$ among its eigenvalues.
Really, a contraction $W$ is the cogenerator of a semigroup $\{W_t\}_{t\in\R^+}$ iff $1$ is not an eigenvalue of $W$; in this case,
$$
W_t=e_t(W)\,,\qquad (t\geq  0)\,,
$$
$$
W=\slim_{t\to0^+} \varphi_t(W_t)\,,
$$
where 
$$
e_t(\lambda):=\exp\left(t{\lambda+1\over \lambda-1}\right)\,,\qquad (t\geq 0)\,,
$$
$$
\varphi_t(\lambda):={\lambda-1+t\over \lambda-1-t}={1-t\over 1+t}-{2t\over 1+t}\sum_{n=1}^\infty {\lambda^n\over (1+t)^n},\qquad (t\geq 0)\,.
$$
Moreover, a semigroup of contractions $\{W_t\}_{t\in\R^+}$ on $\H$ consists of normal, selfadjoint, isometric or unitary operators iff its cogenerator $W$ is normal, selfadjoint, isometric or unitary, respectively. See \cite[Sect.III.8-9]{NAGY-FOIAS}.

Now, given a contraction operator $W$ defined on a Hilbert space $\H$, let $U$ and $U_+$ be the Sz.-Nagy-Foia{\c s} minimal unitary and isometric dilations defined, respectively, on the Hilbert spaces ${\mathcal K}$ and ${\mathcal K}_+$, 
where ${\mathcal K}\supseteq{\mathcal K}_+\supset\H$. 
The {\it residual part} $R$ of $U_+$, defined on the Hilbert space $\RR\subseteq{\mathcal K}_+$, is the unitary part of the Wold decomposition \cite[Th.I.3.2]{NAGY-FOIAS} of $U_+$, i.e.
$$
R:=U_+|\RR\,,\qquad \RR:=\bigcap_{n=0}^\infty U_+^n \H\,.
$$
When $W$ is the cogenerator of a contractive semigroup $\{W_t\}_{t\in\R^+}$ on $\H$, then $U$ and $R$ are the cogenerators of the unitary groups $\{U_t\}_{t\in\R}$ and $\{R_t\}_{t\in\R}$  on ${\mathcal K}$ and $\RR$, respectively, and $U_+$ is the cogenerator of the isometric semigroup $\{U_{+t}\}_{t\in\R^+}$ on ${\mathcal K}_+$, where $U_t$, $U_{+t}$ and $R_t$ are the corresponding minimal unitary and isometric dilations and residual part of $W_t$ for each $t\in\R^+$.

{\it A unitary flow $\{U'_t\}_{t\in\R}$ on ${\mathcal H}'$ is intrinsically irreversible, intertwined with a contraction semigroup $\{W_t\}_{t\in\R^+}$ on ${\mathcal H}$ through a quasi-affinity $\Lambda:\H'\to\H$ satisfying (\ref{Int.Relx}), iff
$\{U'_t\}_{t\in\R}$ is unitarily equivalent to the group of residual parts $\{R_t\}_{t\in\R}$ for $\{W_t\}_{t\in\R^+}$, that is, there exists a unitary operator $V:\H'\to\RR$ such that
\begin{equation}\label{gc.2}
VU'_tV^{-1}=R_t\,,\qquad (t\in\R)\,.
\end{equation}}
\noindent
In such case, 
\begin{equation}\label{gc.1}
\lim_{t\to\infty} W^*_th\neq 0 \text{ for each non-zero }h\in \H
\end{equation}
and the adjoint of the orthogonal projection of ${\mathcal K}$ (or ${\mathcal K}_+$) onto $\RR$ restricted to $\H$, $X:=(P_\RR|\H)^*$, is a quasi-affinity from $\RR$ to $\H$ satisfying  
\begin{equation}\label{gc.3}
W_tX=XR_t\,,\qquad (t\geq0)\,.
\end{equation} 
Thus, $\Lambda=XV$.
Moreover, $V$ can be written in terms of the lifting
$\Lambda _+:\H'\to{\mathcal K}_+$ of $\Lambda$ given by
$$
\Lambda _+=\slim_{n\to\infty} U_+^n\Lambda  U'^{-n}\,.
$$
Indeed,  $\Lambda_+|\Lambda_+|^{-1}$ extends by continuity to $V$, where 
$|\Lambda_+|=(\Lambda_+^*\Lambda_+)^{1/2}$. 

Conversely, for any irreversible flow $\{W_t\}_{t\in\R^+}$ on ${\mathcal H}$, condition (\ref{gc.1}) implies that $X$ is a quasi-affinity from $\RR$ to ${\mathcal H}$ satisfying (\ref{gc.3}), where $\{R_t\}_{t\in\R}$ is the residual group for $\{W_t\}_{t\in\R^+}$, and, moreover, every unitary flow $\{U'_t\}_{t\in\R}$ on ${\mathcal H}'$ verifying the intertwining relation (\ref{Int.Relx}), with $\Lambda$ quasi-affinity, is unitarily equivalent to $\{R_t\}_{t\in\R}$.

The conditions given above in terms of $\{U'_t\}_{t\in\R}$, $\{W_t\}_{t\in\R^+}$ and  $\{R_t\}_{t\in\R}$ can be written in terms of their respective cogenerators $U'$, $W$ and $R$ only. Actually, relations (\ref{Int.Relx})--(\ref{gc.3}) are respectively equivalent to 
$W\Lambda=\Lambda U'$, $VU'V^{-1}=R$,
$\lim_{n\to\infty} W^{*n}h\neq 0$ for each non-zero $h\in {\mathcal H}$ 
and $WX=XR$. This fact permit us to transfer the results from flows to cascades (and vice versa): 

{\it Let $U'$ be a unitary operator on $\H'$ which does not have $1$ as eigenvalue\footnote{
\label{fnote} Since the eigenspace associated to the eigenvalue $1$ of $U$ is just the set of stationary states for which the dynamics $\{U'^n\}_{n\in\Z}$ is trivial, if $1$ is an eigenvalue of $U'$, there is no lost of physical meaning if we consider the cascade 
$\{(U'|\H\ominus\H_1)^n\}_{n\in\Z}$ on $\H\ominus\H_1$, where $\H_1$ is the eigenspace corresponding to the eigenvalue $1$.}.
The unitary cascade $\{U'^n\}_{n\in\Z}$ on ${\mathcal H}'$ is intrinsically irreversible, intertwined with a contraction semigroup $\{W_n\}_{n\in\Z^+}$ on ${\mathcal H}$ through a quasi-affinity $\Lambda:\H'\to\H$ satisfying (\ref{Int.Relx}), iff
$\{U'^n\}_{n\in\Z}$ is unitarily equivalent to the group of residual parts $\{R^n\}_{n\in\Z}$ for $\{W_n\}_{n\in\Z^+}$, that is, there exists a unitary operator $V:\H'\to\RR$ such that
\begin{equation}\label{gc.2c}
VU'^nV^{-1}=R^n\,,\qquad (n\in\Z)\,.
\end{equation}}
\noindent
In such case, $W_n=W^n$ for all $n\in\Z^+$, where $W=W_1$,
\begin{equation}\label{gc.1c}
\lim_{n\to\infty} (W^*)^nh\neq 0 \text{ for each non-zero }h\in \H
\end{equation}
and $X:=(P_\RR|\H)^*$ is a quasi-affinity from $\RR$ to $\H$ satisfying  
\begin{equation}\label{gc.3c}
W^nX=XR^n\,,\qquad (n\in\Z^+)\,.
\end{equation} 
Conversely, for any irreversible cascade $\{W^n\}_{n\in\Z^+}$ on ${\mathcal H}$, condition (\ref{gc.1c}) implies that $X$ is a quasi-affinity from $\RR$ to ${\mathcal H}$ satisfying (\ref{gc.3c}), where $R,\RR$ is the residual part for $W$, and, moreover, for every unitary cascade $\{U'^n\}_{n\in\Z}$ on ${\mathcal H}'$ verifying the intertwining relation (\ref{Int.Relx}), with $\Lambda$ quasi-affinity, $U'$ is unitarily equivalent to $R$.


\section{Functional models}\label{sfm}

Let $\{U'_t\}_{t\in\R}$ [$\{U'^n\}_{n\in\Z}$] be an intrinsically irreversible unitary evolution acting on a Hilbert space ${\mathcal H}'$, connected through the intertwining relation (\ref{Int.Relx}), $\Lambda$ quasi-affinity, with a contraction semigroup $\{W_t\}_{t\in\R^+}$ [$\{W_n\}_{n\in\Z}$] acting on a Hilbert space ${\mathcal H}$.
In Section \ref{sschs} we have seen that $\{U'_t\}_{t\in\R}$ [$\{U'^n\}_{n\in\Z}$] is then unitarily equivalent to the group of residual parts $\{R_t\}_{t\in\R}$ [$\{R^n\}_{n\in\Z}$] of the minimal dilations of $\{W_t\}_{t\in\R^+}$ [$W$]. Therefore, a functional model for $\{R_t\}_{t\in\R}$ and $R$ gives a functional model for $\{U'_t\}_{t\in\R}$ and $U'$. 

In what follows we will pay attention to irreversible systems that approach a unique equilibrium for long times, i.e. to contraction semigroups $\{W_t\}_{t\in\R^+}$ [$\{W_n\}_{n\in\Z}$]  satisfying 
\begin{equation}\label{app.equx}
\slim_{t\to\infty} W_t=0\,, \qquad [\slim_{n\to\infty} W^n=0]\,.
\end{equation}
Here the vectors of $\H$ represent the non-equilibrium deviations of the system in consideration.
In such case, since relations \ref{gc.1} and \ref{gc.1c} are verified as well, we will say, as usual, that $\{W_t\}_{t\geq0}$ and $W$ {\it belong to the class $C_{01}$}. It is well known that a contraction semigroup is of $C_{01}$ class iff its cogenerator also is.  In particular, the contraction operators $W$ and $W_t$ ($t\geq0$) that satisfy (\ref{app.equx}) are completely non unitary (c.n.u.), that is, they lacks of unitary part.

Being the cogenerator $W$ of a semigroup $\{W_t\}_{t\in\R^+}$ of class $C_{01}$ a c.n.u. contraction on $\H$, since $(W^*Wh,h)\leq (h,h)$ and $(WW^*h,h)\leq (h,h)$ for all $h\in\H$, so that $W^*W\leq I_\H$ and $WW^*\leq I_{\H}$, we can define the {\it defect operators}
$$
D_W:=(I_\H-W^*W)^{1/2},\qquad D_{W^*}:=(I_{\H}-WW^*)^{1/2}\,,
$$
which are selfadjoint and bounded by $0$ and $1$, with {\it defect spaces}
$$
{\mathcal D}_W:=\overline{D_W\H}=(\text{Ker }D_W)^\perp\,,\qquad
{\mathcal D}_{W^*}:=\overline{D_{W^*}\H}=(\text{Ker }D_{W^*})^\perp\,.
$$

The {\it characteristic function} of $W$,
$$
\Theta_W(\lambda):=[-W+\lambda D_{W^*}(I-\lambda W^*)^{-1} D_W]_{|{\mathcal D}_W}\,,
$$
is defined at first on the set $A_W$ of all $\lambda\in\C$ such that the operator $I-\lambda W^*$ is boundedly invertible, that is, for $\lambda=0$ and the symmetric image of $\rho(W)\backslash\{0\}$ with respect to the unit circle $C$, where $\rho(W)$ denotes the resolvent set of $W$.
The set $A_W$ is open and contains the open unit disc $D$, and $\Theta_W$ is an analytic function on $A_W$ valued on the set of bounded operators from ${\mathcal D}_W$ into ${\mathcal D}_{W^*}$. In particular $\Theta_W(\lambda)$ is contractive for $\lambda\in D$ and $\Theta_W(\lambda)$ is a unitary operator for $\lambda\in C\cap\rho(W)$.

For almost all $\omega\in C$ the following limit exists
$$
\Theta_W(\omega):=\slim \Theta_W(\lambda)\,,\qquad (\lambda\in D,\,\lambda\to \omega \text{ non-tangentially})\,,
$$
which coincides with the previous definition of $\Theta_W(\omega)$ when $\omega\in A_W$.
Such limits induce a decomposable operator $\Theta_W$ from $L^2({\mathcal D}_W)$ into $L^2({\mathcal D}_{W^*})$ defined by
\begin{equation}\label{chf}
[\Theta_W v](\omega):=\Theta_W(\omega)\,v(\omega)\,,\qquad \text{ for } v\in L^2({\mathcal D}_W)\,.
\end{equation}

For those $\omega\in C$ at which $\Theta_W(\omega)$ exists, thus a.e., set
\begin{equation}\label{df}
\Delta_W(\omega):=[I-\Theta_W(\omega)^*\Theta_W(\omega)]^{1/2}\,.
\end{equation}
$\Delta_W(\omega)$ is a selfadjoint operator on ${\mathcal D}_W$ bounded by $0$ and $1$.
As a function of $\omega$, $\Delta_W(\omega)$ is strongly measurable and generates by
$$
[\Delta_W v](\omega):=\Delta_W(\omega)\,v(\omega)\,,\qquad \text{ for } v\in L^2({\mathcal D}_W)\,,
$$
a selfadjoint operator $\Delta_W$ on $L^2({\mathcal D}_W)$ also bounded by $0$ and $1$.

The functional model for the residual parts $R$ and $\{R_t\}_{t\in\R}$ defined on $\RR$ is described in terms of $\Delta_W$ \cite[Sect.VI.2 and Th.VI.3.1]{NAGY-FOIAS}:
{\it There exists a unitary operator $\Phi$ from $\RR$ onto $\hat \RR$ such that 
$\Phi R=\hat R\Phi$ and $\Phi R_t=\hat R_t\Phi$ ($t\in\R$), where
\begin{equation}\label{fmc}
\left.\begin{array}{ll}
{\hat \RR}:= \overline{\Delta_W L^2({\mathcal D}_W)}\,,
\\[2ex]
{\hat R}(v):= \chi(\omega)\, v(\omega)\,, & (v\in {\hat \RR})\,,
\\[2ex]
{\hat R}_t(v):= e_t(\omega)\, v(\omega)\,, & (v\in {\hat \RR},\,t\in\R)\,.
\end{array}\right\}
\end{equation}
}

Since $\hat R$ is a unitary operator on $\hat \RR$, the subspace {\it $\hat\RR$ is a doubly invariant subspace of $L^2({\mathcal D}_W)$}. 
In particular, if $W$ belongs to the class $C_{01}$, then $\Theta_W(\omega)^*$ is an isometry from ${\mathcal D}_{W^*}$ into ${\mathcal D}_W$ for almost all $\omega\in C$ \cite[Sect.V.2]{NAGY-FOIAS};
the product $\Theta_W(\omega)^*\Theta_W(\omega)$ is then a.e. on $C$ the orthogonal projection from ${\mathcal D}_W$ onto the final subspace of $\Theta_W(\omega)^*$, subspace with dimension equal to that of ${\mathcal D}_{W^*}$; thus, in particular, $\text{dim}\,{\mathcal D}_{W^*}\leq\text{dim}\,{\mathcal D}_W$;
moreover, because the adjoint of a partial isometry is a partial isometry with initial space and final space interchanged \cite[p.63]{HALMOS74}, we have 
$$
\Delta_W(\omega)=P_{\text{Ker}\,\Theta_W(\omega)} \text{ for almost all } \omega\in C\,,
$$
where $P_{\text{Ker}\,\Theta_W(\omega)}$ denotes the orthogonal projection from ${\mathcal D}_{W}$ onto
$\text{Ker}\,\Theta_W(\omega)$; in other words, {\it if $W$ is of class $C_{01}$, then  $\Delta_W$ is a range function and $\hat\RR$ is the corresponding doubly invariant subspace, i.e.
$$
\hat\RR= \M_{\Delta_W}\,.
$$
}

An alternative form to the functional model (\ref{fmc}) can be given in which the roles of the unit disc $D$ and its boundary $C$ are taken over by the complex upper half-plane and the real axis \cite{FO64}.
Indeed, the spaces $L^2({\mathcal D}_W)$ and $H^2({\mathcal D}_W)$ are transformed unitarily --for measures $d\omega/(2\pi)$ on $C$ and $dx/\pi$ on $\R$-- onto, respectively, the space $L^2(\R;{\mathcal D}_W)$ of all (strongly or weakly) Borel measurable functions $f:\R\to {\mathcal D}_W$ such that $\int_\R ||f(x)||^2_{{\mathcal D}_W}\,dx<\infty$ (modulo sets of measure zero) and the Hardy class $H^2(\R;{\mathcal D}_W)$ consisting of the limits on the real axis of the functions $f(z)$ which are analytic on the upper half-plane and for which 
$\sup_{0<y<\infty} \int_\R ||f(x+iy)||_{{\mathcal D}_W}^2\,dx<\infty$.
This is carried out by means of the transformation $u\to f$, where
\begin{equation}\label{crf}
f(x)={1\over x+i}\, u\left({x-i\over x+i}\right),
\end{equation}
and the functional model takes on the following form
\begin{equation}\label{fmr}
\left.\begin{array}{ll}
{\tilde \RR}:= \overline{\Upsilon_W L^2(\R;{\mathcal D}_W)}\,,
\\[1ex]
\ds {\tilde R}(g):= \frac{x-i}{x+i}\, g(x)\,, & (g\in {\tilde \RR})\,,
\\[2ex]
{\tilde R}_t(g):= e^{itx}\, g(x)\,, & (g\in {\tilde \RR},\,t\in\R)\,,
\end{array}\right\}
\end{equation}
with
$$
\Upsilon_W(x):=[I-\Xi_W(x)^*\Xi_W(x)]^{1/2}\,,\qquad
\Xi_W(x):=\Theta_W\left({x-i\over x+i}\right)\,.
$$

The concepts of doubly and simply invariant subspaces and range and rigid operator functions given in Section \ref{sis} can be transfered to the context of the spaces $L^2(\R;\H )$ and $H^2(\R;\H )$: a closed subspace $\M$ of $L^2(\R;\H )$ is called {\it invariant} if $e^{itx}\cdot g(x)\in\M$ for every $g\in\M$ and $t>0$.
$\M$ is called {\it doubly invariant} if $e^{itx}\cdot g(x)\in\M$ for every $g\in\M$ and $t\in\R$.
$\M$ is called {\it simply invariant} if it is invariant but not doubly invariant.
Notice that, for each $t\in\R$, the operator $\tilde R_t$ is just the restriction to ${\tilde \RR}$ of the operator $g\mapsto e^{itx}\cdot g(x)$ defined on $L^2(\R;{\mathcal D}_W)$.
Here the definitions of {\it range function} and {\it rigid operator function} coincide with those given for $L^2(\H)$, but substituting the unit circle $C$ with $\R$.

In the light of the connection --through the transformation (\ref{crf})-- between the functional models given in (\ref{fmc}) and (\ref{fmr}) and since, given a unitary group $\{U_t\}_{t\in\R}$ on $\H$ with cogenerator $U$ and a closed subspace $\M$ of $\H$, the relation $U\M\subseteq \M$ is equivalent to $U_t\M\subseteq\M$ for all $t>0$ (see Lemma II.3.1 in \cite{LP67}), the invariant subspaces of $L^2(\R;\H )$ are characterized as follows:
\begin{itemize}
\item[(1)]
{\it The doubly invariant subspaces of $L^2(\R;\H )$ are the subspaces $\M_J$, where $J$ is a measurable range function on $\R$.} The correspondence between $J$ and $\M_J$ is one-to-one, under the convention that range functions are identified if they are equal a.e. on $\R$.
\item[(2)]
{\it Each simply invariant subspace $\M$ of $L^2(\R;\H )$ has the form
$$
\tilde U\, H^2(\R;\H)\oplus \M_K\,,
$$
where $K$ is a measurable range function on $C$ and $\tilde U$ is a rigid ${\mathcal L}(\H )$-valued function on $\R$ with range $J$ orthogonal to $K$ almost everywhere.} The subspace $\tilde U\, H^2(\R;\H)$ uniquely determines the rigid operator function $\tilde U$ to within a constant partially isometric factor on the right.
\end{itemize}

Using similar arguments as before, it is possible to show that {\it $\Upsilon_W$ is a range function on $\R$ and $\tilde\RR$ is the corresponding doubly invariant subspace, i.e.
$$
\tilde\RR= \M_{\Upsilon_W}\,.
$$
}


\section{Selfadjoint  time operators}\label{sito}

For continuous time parameter we have the following set of equivalent conditions to the existence of a selfadjoint  time operator for a unitary flow:
 
\begin{itemize}
\item[I.]
{\it For a unitary flow $\{U_t\}_{t\in\R}$ defined on a separable Hilbert space $\H$ the following assertions are equivalent:}
\begin{itemize}
\item[(I.1)]
there exists a {\it selfadjoint  time operator} $T$ for $\{U_t\}_{t\in\R}$ satisfying the WWR; 

\item[(I.2)]
$\{U_t\}_{t\in\R}$ is {\it intrinsically irreversible}, so that it is unitarily equivalent to the functional model $\{\tilde R_t\}_{t\in\R}$ on $\tilde \RR$ given in (\ref{fmr}), and, moreover, there exists a {\it rigid ${\mathcal L}({\mathcal D}_W)$-valued function} $\tilde U$ on $\R$ with range $\tilde\RR$, i.e. such that
\begin{equation}\label{sb2xrf2}
\overline{\bigcup_{t\in\R}\tilde R_t\,\tilde U H^2({\mathcal D}_W)}=\tilde\RR\,;
\end{equation}

\item[(I.3)]
$\{U_t\}_{t\in\R}$ is {\it intrinsically irreversible} with functional model $\{\tilde R_t\}_{t\in\R}$ on $\tilde \RR$ given in (\ref{fmr}) and, moreover,
there exists a {\it simply invariant subspace} $\tilde\M\subseteq\tilde\RR$ of $L^2(\R;{\mathcal D}_W)$ such that 
\begin{equation}\label{sb2xi2}
\bigcap_{t\in\R} \tilde R_t\,\tilde\M=\{0\}\,,\quad
\overline{\bigcup_{t\in\R}\tilde R_t\,\tilde\M}=\tilde\RR\,;
\end{equation}

\item[(I.4)]
there exists an {\it outgoing subspace} for $\{U_t\}_{t\in\R}$, i.e. a closed subspace $\M_+\subseteq\H$ with the following properties:
$$
U_t\M_+\subseteq \M_+\,, \quad \text{for all } t\in\R^+\,,
$$
$$
\bigcap_{t\in\R}\, U_t\,\M_+=\{0\}\,,\quad
\overline{\bigcup_{t\in\R}\,U_t\,\M_+}=\H\,;
$$
equivalently, there exists an {\it incoming subspace} $\M_-\subseteq\H$ for $\{U_t\}_{t\in\R}$ (see below);

\item[(I.5)]
$\{U_t\}_{t\in\R}$ is a (continuous-parameter) {\it purely nondeterministic innovation process} (see below),
here the innovation subspaces are defined to be 
$$
{\mathcal I}_{t_2,t_1}=U_{t_2}{\mathcal M_-}\ominus U_{t_1}{\mathcal M_-}\,, 
\quad \text{for }\,\, t_1\leq t_2\,;
$$ 

\item[(I.6)]
$\{U_t\}_{t\in\R}$ is a (continuous-parameter) {\it Lax--Phillips scattering process} (see below);

\item[(I.7)]
{\it translation representation:} $\H$ can be represented as $L^2(\R;{\mathcal N})$, where ${\mathcal N}$ is an auxiliary Hilbert space, in such a way that $U_t$ corresponds to $u_t$, the right translation by $t$ units defined by 
\begin{equation}\label{bstr}
[u_tg](x):= g(x-t)\,,\quad (g\in L^2(\R;{\mathcal N}))\,,
\end{equation} 
$T$ to the operator $q$ multiplication by the independent variable, and $\M_+$ maps onto the subspace $L^2(\R^+;{\mathcal N})$ of $L^2(\R;{\mathcal N})$; 

\item[(I.8)]
{\it spectral representation:} $\H$ can be represented as $L^2(\R;{\mathcal N})$, where ${\mathcal N}$ is an auxiliary Hilbert space, so that $U_t$ goes into multiplication by $e^{ixt}$, i.e. 
$$
g\mapsto e^{ixt}g(x)\,,\quad (g\in L^2(\R;{\mathcal N});\, x,t\in\R)\,,
$$ 
$T$ into $p=-id/dx$, and $\M_+$ is mapped onto $H^2(\R;{\mathcal N})$, the Hardy class of functions $g(x)$ whose Fourier transform vanishes for all negative $x$. 

\item[(I.9)]
there exists a unitary group $\{V_t\}_{t\in\R}$ on $\H$ such that the {\it Weyl commutation relation}
\begin{equation}\label{h702ax}
U_t\, V_s=e^{its}\,V_s\,U_t\,,\qquad (t,s\in\R)\,,
\end{equation}
is satisfied; in fact, $V_t=e^{itT}$, $(t\in\R)$, where $T$ is a time operator for $\{U_t\}_{t\in\R}$;

\item[(I.10)]
if $A$ be the selfadjoint operator such that $U_t=e^{itA}$, $(t\in\R)$, there exists a selfadjoint operator $T$ on $\H$ such that $(A,T)$ is a {\it Schr\"odinger couple} (see below) or a direct sum of such couples; $T$ is then a time operator for $\{U_t\}_{t\in\R}$. 

\end{itemize}
\end{itemize}

As a direct corollary of (I.10) we have that for a unitary flow $\{U_t=e^{itA}\}_{t\in\R}$ with  time operator $T$ the spectra $\sigma(T)$ and $\sigma(A)$ satisfy
$$
\sigma(T)=\sigma_{ac}(T)=\sigma(A)=\sigma_{ac}(A)=\R
$$
($\sigma_{ac}$ denotes the absolutely continuous spectrum with respect to Lebesgue measure) and have the same constant multiplicity.

\medskip

For discrete time parameter the following set of equivalent conditions to the existence of a selfadjoint  time operator for a unitary cascade can be deduced:

\begin{itemize}

\item[II.]
{\it If $\{U^n\}_{n\in\Z}$ is a unitary cascade defined on a separable Hilbert space $\H$ such that $1$ is not an eigenvalue of the unitary operator $U$\footnote{See footnote on page \pageref{fnote}.}, the following assertions are equivalent:}
\begin{itemize}
\item[(II.1)]
there exists a {\it selfadjoint  time operator} $T$ for $\{U^n\}_{n\in\Z}$ satisfying the WWR; 

\item[(II.2)]
$\{U^n\}_{n\in\Z}$ is {\it intrinsically irreversible}, so that it is unitarily equivalent to the functional model $\{\hat R^n\}_{n\in\Z}$ on $\hat \RR$ given in (\ref{fmc}), and, moreover, there exists a {\it rigid ${\mathcal L}({\mathcal D}_W)$-valued function} $\hat U$ on $C$ with range $\hat\RR$, i.e. such that
\begin{equation}\label{sb2xrf1}
\overline{\bigcup_{n\in\Z}\hat R^n\,\hat U H^2({\mathcal D}_W)}=\hat\RR\,;
\end{equation}

\item[(II.3)]
$\{U^n\}_{n\in\Z}$ is {\it intrinsically irreversible} with functional model $\{\hat R^n\}_{n\in\Z}$ on $\hat \RR$ given in (\ref{fmc}) and, moreover,
there exists a {\it simply invariant subspace} $\hat\M\subseteq\hat\RR$ of $L^2({\mathcal D}_W)$ such that 
\begin{equation}\label{sb2xi1}
\bigcap_{n\in\Z} \hat R^n\,\hat\M=\{0\}\,,\quad
\overline{\bigcup_{n\in\Z}\hat R^n\,\hat\M}=\hat\RR\,;
\end{equation}

\item[(II.4)]
there exists an {\it outgoing subspace} for $\{U^n\}_{n\in\Z}$, i.e. a closed subspace $\M_+\subseteq\H$ with the following properties:
$$
U^n\M_+\subseteq \M_+\,, \quad \text{for all } n\in\Z^+\,,
$$
$$
\bigcap_{n\in\Z}\, U^n\,\M_+=\{0\}\,,\quad
\overline{\bigcup_{n\in\Z}\,U^n\,\M_+}=\H\,;
$$
equivalently, there exists an {\it incoming subspace} $\M_-\subseteq\H$ for $\{U^n\}_{n\in\Z}$ (see below);

\item[(II.5)]
$\{U^n\}_{n\in\Z}$ is a (discrete-parameter) {\it purely nondeterministic innovation process} (see below),
here the innovation subspaces are defined to be 
$$
{\mathcal I}_{n_2,n_1}=U^{n_2}{\mathcal M_-}\ominus U^{n_1}{\mathcal M_-}\,, 
\quad \text{for }\,\, n_1\leq n_2\,;
$$ 

\item[(II.6)]
$\{U^n\}_{n\in\Z}$ is a (discrete-parameter) {\it Lax--Phillips scattering process} (see below);

\item[(II.7)]
{\it translation representation:} $\H$ can be represented as $l^2(\Z;\NN)$ for some auxiliary Hilbert space $\NN$, so that $U$ goes into the right shift operator, $T$ into the multiplication operator by the independent variable, and $\M_+$ maps onto $l^2(\Z^+;\NN)$;

\item[(II.8)]
{\it spectral representation:} 
$\H$ can be represented as $L^2(\NN)$ for some auxiliary Hilbert space $\NN$ so that $U$ goes into multiplication by $\om$, $T$ into $p=-i\,d/d\theta$, $\theta$ the polar angle, and $\M_+$  maps onto $H^2(\NN)$, the Hardy class of functions $f(\om)$ whose kth Fourier coefficients vanish for all negative k's.

\end{itemize}

\end{itemize}

Moreover, {\it when we consider a unitary flow $\{U_t\}_{t\in\R}$ and its unitary cogenerator $U$, the equivalent conditions (I.1)--(I.10) for $\{U_t\}_{t\in\R}$ are, at the same time, equivalent to the equivalent conditions (II.1)--(II.8) for the unitary cascade $\{U^n\}_{n\in\Z}$.}
This is due to the fact that, for a closed subspace $\M$, the relation $U\M\subseteq \M$ is equivalent to $U_t\M\subseteq\M$ for all $t>0$ and, if any of both relations is satisfied (and then both of them), then   
$$
\bigcap_{t\in\R} U_t\,\M=\bigcap_{n\in\Z} U^n\,\M\,,\qquad
\overline{\bigcup_{t\in\R}U_t\,\M}=\overline{\bigcup_{n\in\Z}U^n\,\M}\,,
$$
(see Lemmas II.3.1--2 in \cite{LP67}).

\medskip

To prove the equivalences 
(I.1)$\Leftrightarrow$(I.2)$\Leftrightarrow$(I.3)$\Leftrightarrow$(I.9) 
let us begin by 
assuming that $\{U_t\}_{t\in\R}$ is intrinsically irreversible, so that it is unitarily equivalent to the functional model $\{\tilde R_t\}_{t\in\R}$ on $\tilde \RR$ given in (\ref{fmr}), and that there exists a rigid ${\mathcal L}({\mathcal D}_{W} )$-valued function $\tilde U$ on $\R$ with range ${\tilde \RR}$, i.e. a weakly measurable ${\mathcal L}({\mathcal D}_{W})$-valued function $\tilde U$ on $\R$ such that $\tilde U(x)$ is for almost all $x\in \R$ a partial isometry on ${\mathcal D}_{W}$ with the same initial space and
\begin{equation}\label{kc}
\bigcap_{t<\infty} {\tilde R}_t\,\tilde\M=\{0\}\,,\qquad
\overline{\bigcup_{t>-\infty}{\tilde R}_t\,\tilde\M}={\tilde \RR}\,,
\end{equation}
where $\tilde\M:=\tilde U\,H^2(\R;{\mathcal D}_W)$. This is feasible since ${\tilde \RR}$ is a doubly invariant subspace. See Sections \ref{sis} and \ref{sfm}.
By the Paley-Wiener representation of $H^2(\R;{\mathcal D}_W)$ \cite[Th.4.8.E]{RR85}, 
$$
{\tilde R}_t\,H^2(\R;{\mathcal D}_W) \subseteq {\tilde R}_{t'}\,H^2(\R;{\mathcal D}_W)\,, \qquad \text{if }\,\, t>t'\,,
$$
and, since $\tilde U$ commutes with $\tilde R_t$ for each real $t$, we have
\begin{equation}\label{kf}
{\tilde R}_t\,\tilde\M \subseteq {\tilde R}_{t'}\,\tilde\M\,, \qquad \text{if }\,\, t>t'\,.
\end{equation}
Thus, if $\tilde P_t$ denotes the orthogonal projection of ${\tilde \RR}$ onto ${\tilde R}_t\,\tilde\M$ for each real $t$, by virtue of (\ref{kc}) and (\ref{kf}),
the family $\{I-\tilde P_t\}$ is a decomposition of the identity in  ${\tilde \RR}$ and we can construct the unitary group
\begin{equation}\label{gto}
\tilde V_s:=-\int_\R \exp(its)\,d\tilde P_t
\end{equation}
with selfadjoint generator
\begin{equation}\label{to}
\tilde T:=-\int_\R t\,d\tilde P_t\,.
\end{equation}
From definition, $\tilde P_{t+s}={\tilde R}_t\,\tilde P_s\,{\tilde R}_{-t}$ for all $t,s\in\R$; indeed, being $\{{\tilde R}_t\}_{t\in\mathbb R}$ a unitary group, ${\tilde R}_t\,\tilde P_s\,{\tilde R}_{-t}$ is the orthogonal projection onto ${\tilde R}_t\,{\tilde R}_s\,\tilde\M={\tilde R}_{t+s}\,\tilde\M$.
In consequence, we have the two commutation relations
\begin{equation}\label{h702a}
\tilde V_s\,{\tilde R}_t=\exp(its)\,{\tilde R}_t\,\tilde V_s\,,\qquad (t,s\in\R)\,,
\end{equation}
\begin{equation}\label{h702b}
{\tilde R}_{-t}\,\tilde T\,{\tilde R}_t=\tilde T+t\tilde I\,,\qquad (t,s\in\R)\,,
\end{equation}
where $\tilde I$ denotes the identity in ${\tilde \RR}$. Relation (\ref{h702b}) means, in particular, that the domain of $\tilde T$ is invariant under all ${\tilde R}_t$ and that {\it $\tilde T$ is an  time operator for the unitary group $\{\tilde R_t\}_{t\in\R}$.}

Reciprocally, each continuous unitary group $\{\tilde V_s\}$ acting in ${\tilde \RR}$ and satisfying (\ref{h702a}) --or selfadjoint operator $\tilde T$ satisfying (\ref{h702b})-- leads backwards to a unique subspace $\tilde\M$ and almost unique rigid function $\tilde U$, save a constant partially isometric factor on the right. 

Taking into account that every unitary dynamics with time operator is intrinsically irreversible, being the quasi-affinity $\Lambda$ an operator function of the time operator and the irreversible dynamics converging to a unique equilibrium for long times \cite{M78,S92}, coming back to the original unitary evolution and Hilbert space we obtain the equivalences (I.1)$\Leftrightarrow$(I.2)$\Leftrightarrow$(I.3)$\Leftrightarrow$(I.9).

The equivalence (I.9)$\Leftrightarrow$(I.10) is due to von Neumann \cite{VN31} (see also \cite[Th.VIII.14]{RS72}), who proved that two unitary groups $\{V_t\}_{t\in\R}$ and $\{U_t\}_{t\in\R}$ defined on a separable Hilbert space satisfy the Weyl commutation relation (\ref{h702ax}) iff their respective generators $T$ and $A$ form a Schr\"odinger couple or a direct sum of such couples, where by a {\it Schr\"odinger couple} we mean the pair $(P,Q)$ on $L^2(\R;\C)$ defined by 
\begin{equation}\label{mpo}
[Pg](x):=-i\,g'(x)\quad \text{and} \quad [Qg](x):=x\,g(x)\,,
\end{equation}
the domains given by (AC:=absolutely continuous)
\begin{equation}\label{mpod}
\begin{array}{l}
\text{Dom}(P)=\{g\in L^2(\R;\C): g\text{ AC and } g'\in L^2(\R;\C)\}\,,
\\
\text{Dom}(Q)=\{g\in L^2(\R;\C):  x\,g(x)\in L^2(\R;\C)\}\,,
\end{array}
\end{equation}
or any other pair of operators unitarily equivalent to this one.

Suchanecki and Weron \cite{SW90} obtained the equivalences (I.1)$\Leftrightarrow$(I.9)$\Leftrightarrow$(I.10) restricting attention to Kolmogorov flows. 

Clearly, if we transfer the invariant subspace $\hat\M$ from $\hat\R$ into the space where the  unitary evolution $\{U_t\}_{t\in\R}$ lives, we obtain an outgoing subspace $\M_+$, and reciprocally.
Incoming subspaces are defined similarly to outgoing ones. An {\it incoming subspace} for $\{U_t\}_{t\in\R}$ is a closed subspace $\M_-\subseteq\H$ with the following properties:
\begin{equation}\label{isb1x}
U_t\M_-\subseteq \M_-\,, \quad \text{for all } t\in\R^-:=\R\backslash\R^+\,,
\end{equation}
\begin{equation}\label{isb2x}
\bigcap_{t\in\R}\, U_t\,\M_-=\{0\}\,,\quad
\overline{\bigcup_{t\in\R}\,U_t\,\M_-}=\H\,.
\end{equation}
Note that if $\M_-$ [$\M_+$] is incoming [outgoing], then the orthogonal complement of $\M_-$ [$\M_+$] is outgoing [incoming].
Taking into account (\ref{kf}), these facts lead to the equivalence (I.3)$\Leftrightarrow$(I.4).

Furthermore, whenever there is an outgoing [incoming] subspace for the group $\{U_t\}_{t\in\R}$ there is always an {\it outgoing [incoming] translation representation}.  
This result was obtained by {Sina\u\i} \cite{Sinai61} (see also \cite[Prop.III.9.4]{NAGY-FOIAS}).  For outgoing subspaces it reads as follows: 
Every group $\{U_t\}_{t\in\R}$ of unitary operators acting on a Hilbert space $\H$, for which there is an outgoing subspace $\M_+$, is unitarily equivalent to the bilateral shift $\{u_t\}_{t\in\R}$ on $L^2(\R;{\mathcal N})$, defined by (\ref{bstr}),
where ${\mathcal N}$ is any Hilbert space whose dimension equals that of $\M_+\ominus U^{(r)}\M_+$, $U^{(r)}$ denoting the cogenerator of the semigroup formed by the restrictions
$$
U^{(r)}_t:=U_{t}|\M_+\,,\quad (t\in\R^+)\,.
$$
Moreover, the unitary operator effectuating this unitary equivalence can be chosen so that it maps $\M_+$ onto the subspace $L^2(\R^+;{\mathcal N})$ of $L^2(\R;{\mathcal N})$. Therefore, from {Sina\u\i} theorem, (I.4)$\Rightarrow$(I.7). 

That (I.7)$\Rightarrow$(I.1) is obvious since, for all $g\in\text{Dom}(q)$ and all $t\in\R$, we have $u_tg\in\text{Dom}(q)$ and
$$
[u_{-t}qu_tg](x)=xg(x)+tg(x)\,.
$$
 
To prove the equivalence (I.7)$\Leftrightarrow$(I.8) it is enough to consider the Fourier transformation $F$ and its inverse to obtain the spectral representation of (I.7) from the translation representation of (I.8) and vice versa. According to the Paley-Wiener theorem, $FL^2(\R^+;{\mathcal N})=H^2(\R;{\mathcal N})$. See \cite[Sect.4.8]{RR85} and \cite[Sect.II.3]{LP67} for details.

The existence of outgoing and incoming subspaces $\M_+$ and $\M_-$ characterizes the unitary evolutions the {\it Lax-Phillips scattering theory} \cite{LP67} deals with. Therefore, (I.4)$\Leftrightarrow$(I.6).
In Lax-Phillips scattering theory distinguished translation and spectral representations, as those of items (I.7) and (I.8), are associated to the different outgoing and incoming subspaces, so that representations derived from mutually orthogonal subspaces $\M_+$ and $\M_-$ are related by means of a scattering operator $S$. In some sense, $S$ connects the behavior in the remote past with that in the distant future.
The scattering operator $S$ can be realized as an operator-valued function $\mathcal S$. The function $\mathcal S$ is just the characteristic function of the cogenerator of the contraction semigroup $\{Z_t\}_{t\in\R^+}$ constructed from $\{U_t\}_{t\in\R}$ by means of a truncation process into the orthogonal complement of the outgoing and incoming subspaces $\M_+\oplus\M_-$. 
On the other hand, due to the intrinsic irreversibility, $\{U_t\}_{t\in\R}$ is completely determined by the contraction semigroup $\{W_t\}_{t\in\R^+}$ and the characteristic function $\Theta_W$ of its cogenerator through the functional models given in Section \ref{sfm}.
At first sight the semigroup $\{Z_t\}_{t\in\R^+}$ and the scattering function $\mathcal S$
have nothing to do with $\{W_t\}_{t\in\R^+}$ and $\Theta_W$ --among other things, $\mathcal S$ is unitary on the boundary, whereas $\Theta_W$ is not \cite{G}--, but this question, as well as the role that the rigid function $\tilde U$ of (I.2) (or corresponding $\hat U$) plays in this framework, deserve further study.

Now, let us recall that in the least squares prediction theory for stochastic processes \cite[Chapter XII]{D53} a process is associated to the increasing family $\{\M_t\}_{t\in\R}$ of closed subspaces generated by the random variables of the process up to time $t$ or their corresponding spectral representations; actually, $\{\M_t\}_{t\in\R}$ is determined by $\M_0$ \cite[p. 587]{D53}. Moreover, the process lacks of deterministic part iff $\{\M_t\}_{t\in\R}$ is strictly increasing or, equivalently,
\begin{equation}\label{pndsp}
\bigcap_{t\in\R}\, \M_t=\{0\}\,;
\end{equation}
in this case, the process is called {\it regular} or {\it purely nondeterministic}.

In a similar way, by an {\it innovative unitary evolution} we mean a unitary group $\{U_t\}_{t\in\R}$ for which there exists an increasing family of closed subspaces $\{\M_t\}_{t\in\R}$ such that
$$
U_{t_1}\M_{t_2}=U_{t_2}\M_{t_1}\,,\qquad (t_1,t_2\in\R)\,.
$$
The innovative evolution will be called {\it purely nondeterministic} if (\ref{pndsp}) is satisfied.
The concept of purely nondeterministic innovation evolution can be considered a generalization of the concept of {\it Kolmogorov flow}, i.e. an abstract dynamical system $(\Omega,{\mathcal A},\mu,\{S_t\})$ for which there exists a sub $\sigma$-algebra ${\mathcal A}_0\subset{\mathcal A}$ with the following properties \cite{LM94}:
$$
{\mathcal A}_t:=S_t{\mathcal A}_0 \subseteq {\mathcal A}_{t'}\,, \qquad \text{if }\,\, t<t'\,,
$$
$$
\bigcap_{t<\infty} {\mathcal A}_t=\text{trivial $\sigma$-algebra}\,,\qquad
\sigma\Big(\bigcup_{t>-\infty}{\mathcal A}_t\Big)={\mathcal A}\,.
$$
In the light of the proof of the equivalences (I.1)$\Leftrightarrow$(I.3)$\Leftrightarrow$(I.4) given above, every innovative unitary evolution $\{U_t\}_{t\in\R}$ is qualified by the existence of an  time operator or, equivalently, by the existence of an incoming subspace $\M_-=\M_0$, and conversely. This completes the proof.


\section{The Aharonov-Bohm time operator}\label{sabto}

Aharonov and Bohm \cite{AB61} considered the symmetric time operator 
$$
T_0=\frac12\left(QP^{-1}+P^{-1}Q\right)\,,
$$ 
for the one-dimensional free non-relativistic spinless Hamiltonian $H_0=P^2/2$ on the Hilbert space $L^2(\R;\C)$, where the momentum operator $P$ and the position operator $Q$ are defined by (\ref{mpo}) and (\ref{mpod}) (for $\hbar=1$ and inertial mass $m=1$). Since $P$ is injective, the inverse $P^{-1}$ is well defined and becomes a selfadjoint operator on $L^2(\R;\C)$. The domain of $T_0$ is $\text{Dom}(T_0)=\text{Dom}(QP^{-1})\cap\text{Dom}(P^{-1}Q)$.

Taking Fourier transforms, the momentum representation of $T_0$ is
$$
\begin{array}{l}
\ds FT_0F^{-1}=\frac{i}{2}\Big(\frac{d}{dk}\,M_{1/k}+M_{1/k}\,\frac{d}{dk}\Big)\,,\\[1ex]
\ds \text{Dom}(FT_0F^{-1})=\text{Dom}\Big(\frac{d}{dk}\,M_{1/k}\Big)\cap\text{Dom}\Big(M_{1/k}\,\frac{d}{dk}\Big)\,,
\end{array}
$$
where $M_{v(k)}$ denotes the operator multiplication by $v(k)$. The operator $T_0$ does not satisfy the WWR for the time evolution group $U_t=e^{-itH_0}$ because $\text{Dom}(FT_0F^{-1})$ is not invariant under the action of $e^{-itM_{k^2/2}}$ for all $t\neq 0$. However the following symmetric extension $\tilde T_0$ of $T_0$ does (see \cite{Mi01}):
\begin{equation}\label{fto}
\begin{array}{l}
\ds F\tilde T_0F^{-1}\,\hat\psi(k)=\frac{i}{2}\Big(\frac{d\hat\psi(k)/k}{dk}+\frac{1}{k}\frac{d\hat\psi(k)}{dk}\Big)\,, \text{ for a.e. }k\in\R\backslash\{0\}\,,\\[1ex]
\ds 
\text{Dom}(F\tilde T_0F^{-1})=
\\
=\left\{\hat\psi\in L^2(\R;\C)\left|
\begin{array}{l}
\ds \hat\psi\in \text{AC}(\R\backslash\{0\}),\, \lim_{k\to 0} \frac{\hat\psi(k)}{|k|^{1/2}}=0,\text{ and}\\
\ds \int_{\R\backslash\{0\}}\left| \frac{d\hat\psi(k)/k}{dk}+\frac{1}{k}\frac{d\hat\psi(k)}{dk}\right|^2\,dk<\infty
\end{array}
\right.\right\}\,,
\end{array}
\end{equation}
where $\hat\psi=F\psi$ denotes the Fourier transform of $\psi$.
The operator $\tilde T_0$ is interpreted as a {\it time-of-arrival operator} (see \cite{EM00} and references therein).

On the other hand, the momentum representation of $H_0$ is given by
\begin{equation}\label{fho}
\begin{array}{l}
\ds FH_0F^{-1}\,\hat\psi(k)=\frac{k^2}{2}\,\hat\psi(k)\,,\\
\ds \text{Dom}(FH_0F^{-1})=\big\{\hat\psi\in L^2(\R;\C): 
\int_\R |k^2\, \hat\psi(k)|^2\,dk<\infty\big\}\,,
\end{array}
\end{equation}
so that, to obtain the {\it energy representation}, useful in scattering theory,
it is enough to set $\l=k^2/2$ and, for each $\psi\in L^2(\R;\C)$ and for almost $\l\in
[0,\infty)$, the vector $\vec\psi(\l)\in \C^2$ given by
$$
\vec\psi(\l):= \frac{1}{(2\l)^{1/4}}\Big( \hat\psi(+\sqrt{2\l}),\hat\psi(-\sqrt{2\l})\Big)\,,
$$
in order to have
$||\psi||^2=||\hat \psi||^2 =\int_0^\infty ||\vec\psi(\l)||^2_{\C^2}\,d\l$.
Thus, $L^2(\R;\C)$ is unitarily isomorphic to $L^2\big([0,\infty);\C^2)$, the isomorphism $V$ being given by $V\psi=\vec\psi$. After straightforward calculations, from (\ref{fto}) and (\ref{fho}) we get
\begin{equation}\label{eto}
\begin{array}{l}
\ds V\tilde T_0V^{-1}\,\vec\psi(\l)=-i\,\vec\psi'(\l)\,,\quad \text{for a.e. }\l\in(0,\infty)\,,\\[1ex]
\ds 
\text{Dom}(V\tilde T_0V^{-1})=
\\
=\left\{\vec\psi\in L^2([0,\infty);\C^2)\left|
\begin{array}{l}
\ds \vec\psi\in \text{AC}(0,\infty),\, \lim_{\l\to 0} \vec\psi(\l)=0,\\
\ds \vec\psi' \in L^2([0,\infty);\C^2)
\end{array}
\right.\right\}\,,
\end{array}
\end{equation}
and
\begin{equation}\label{eho}
\begin{array}{l}
\ds VH_0V^{-1}\,\tilde\psi(\l)=\l\,\vec\psi(\l)\,,\\
\ds \text{Dom}(FH_0F^{-1})=
\\
=\big\{\tilde\psi\in L^2([0,\infty);\C^2): 
\int_0^\infty ||\l\,\vec\psi(\l)||_{\C^2}^2\,d\l<\infty\big\}\,,
\end{array}
\end{equation}

It is clear that extending the physical Hamiltonian $VH_0V^{-1}$ and its domain to negative energies by $VH_0V^{-1}\oplus(-VH_0V^{-1})$ we obtain the usual selfadjoint operator $Q$ over $L^2(\R;\C^2)$. A similar extension of $V\tilde T_0V^{-1}=-i\,d/d\l$ does not give a selfadjoint operator due to the additional condition $\lim_{\l\to 0} \vec\psi(\l)=0$ in its domain. To see this remind that a symmetric operator $T$ on a Hilbert space $\H$ is selfadjoint iff $\text{Rang}(T\pm i)=\H$; now, if we consider $T=-i\,d/d\l$ with domain $\text{Dom}(T)=\{f\in L^2(\R;\C): f\in\text{AC},\,\lim_{\l\to 0} f(\l)=0,\,f'\in L^2(\R;\C)\}$, we have $g\in\text{Rang}(T\pm i)$ iff there exists $f\in\text{Dom}(T)$ such that   
$$
f(\l)= i\,\exp(\pm \l)\int_0^\l \exp(\mp s)\,g(s)\,ds\,,
$$
and obviuosly not every $g\in L^2(\R;\C)$ satisfies this condition. According to the equivalence 
(I.1)$\Leftrightarrow$(I.10) of Section \ref{sito} we must drop the additional condition from the domain definition in order to obtain the required selfadjoint extension of $V\tilde T_0V^{-1}$ as the twofold usual momentum operator $P$.
The selfadjoint operator $V^{-1}PV$ is just the Werner dilation \cite{Wer90} of the Aharonov-Bohm time operator $\tilde T_0$. Correspondingly, the projection valued measure associated to $V^{-1}PV$ is the Naimark dilation of the positive operator valued measure determined by $\tilde T_0$ \cite{Wer90,EM00}.

The Fourier transform of the extended energy representation leads to the {\it time representation} where the probability density for the time of arrival over a pure state is nothing but the modulus squared of the wavefunction. The usual probability density for time of arrival is reobtained by restriction to the initial Hilbert space \cite{EM00}.

This procedure of extending physical Hamiltonians $H$ by $H\oplus(-H)$ can also be considered in order to connect quantum mechanical evolutions and two-space scattering with the Lax-Phillips framework. It tuns out, for example, that the primary form of the Gamov vectors can be derived using the spectral representation of the extended Hamiltonian. Besides, the imposibility to obtain outgoing and incoming subspaces simultaneously with respect to a reference Lax-Phillips process indicates the existence of a time arrow \cite{Baum03}.


\section*{Acknowledgments}

The author wishes to thank Profs. M. N\'u\~nez and Z. Suchanecki for useful discussions and the staff of {\it Facult\'e des Sciences, de la Technologie et de la Communication} of {\it Universit\'e du Luxembourg} for kind hospitality. 
This work was supported by JCyL-project VA013C05 (Castilla y Le\'on) and MEC-project FIS2005-03989 (Spain).

\bigskip


\end{document}